\documentclass[pre,floatfix,twocolumn, 10pt]{revtex4-1} 
\usepackage{graphicx}
\usepackage{amssymb}
\usepackage{natbib}
\usepackage{dcolumn}
\usepackage{bm}
\usepackage{color}
\usepackage[colorlinks=true, linkcolor=blue, citecolor=blue]{hyperref}
\usepackage{subfigure}
\usepackage{graphicx,psfrag,xspace}
\usepackage{color}
\usepackage{amssymb,amsfonts,amsmath}
\usepackage{setspace}

\begin{document}


\author{E. A. Jagla} 
\affiliation{Centro At\'omico Bariloche, Instituto Balseiro, 
Comisi\'on Nacional de Energ\'ia At\'omica, CNEA, CONICET, UNCUYO,\\
Av.~E.~Bustillo 9500 (R8402AGP) San Carlos de Bariloche, R\'io Negro, Argentina}

\title{Modeling plasticity of amorphous composites: Scalar is not enough}
\begin{abstract} 

We use a continuous mesoscopic model to address the yielding properties of plastic composites, formed by a host material and
inclusions with different elastic and/or plastic properties. We investigate the flow properties of the composed material under a uniform externally applied  deviatoric stress. We show that due to the heterogeneities induced by the inclusions, a scalar modeling in terms of a single deviatoric strain of the same symmetry than the externally applied deformation gives inaccurate results. 
A realistic modeling must include all possible shear deformations. Implementing this model in a two-dimensional system we show that the effect of harder inclusions is very weak up to relatively high concentrations. For 
softer inclusions instead, the effect is much stronger, even a small concentration of inclusions affecting the form of the flow curve and the critical stress. We also present the details of a full three dimensional simulation scheme, and obtain the corresponding results, both for harder and softer inclusions.

\end{abstract}

\maketitle

\section{Introduction}

The properties of heterogeneous materials have been the subject of study for many years. From a practical point of view, in many cases the addition of a fraction of foreign particles with different properties than the host has been widely used as a mean to improve the mechanical properties
of the sample\cite{torquato}. In the case of plastic materials, there is a finite shear strength that is macroscopically supported by the sample and can be taken as an example of a mechanical property that can be maximized.

From the theoretical point of view, the study of the properties of plastic materials has had an important advance in the last years\cite{vandem_review}. This has been boosted by the importance that materials of this kind have gained both in industrial processes and every day life\cite{rmp2}.
In this respect, many remarkable findings have been obtained through the use of numerical techniques. Besides other possibilities, for our present purposes we want to concentrate here in a modeling method that has been termed Elasto-Plastic Modeling\cite{rmp,talamali,pnas,zapperi,aguirre}. In this scheme, the system is represented in a coarse grained manner over an ordered (usually cubic) lattice, with each site representing a small portion of the real material. Different elements in the system interact through elastic couplings that are defined by the usual elasticity properties of the material. The possibility of plastic rearrangements are contained in the internal dynamics of each mesoscopic element in the system.

It is clear that a full description of the elasto-plastic behavior of materials requires a full description of elasticity, and thus a consideration of the full strain elastic tensor. This is mandatory in cases in which macroscopic inhomogeneous experiments or samples are being studied. Indentation is a typical case\cite{indent}: the directions of the principal axis of the strain tensor vary across the sample, and thus the consideration of a single shear deformation is clearly insufficient.

However, in many cases one is interested in how the microscopic properties that are fed into the model manifest in the macroscopic behavior of the material (such as its yielding behavior), and in this case the simplest case of a homogeneous sample (numerically accomplished by the use of periodic boundary conditions) under a single mode external shear stress is the situation we are interested in. If this is the case, it has been argued that the full tensorial description can be simplified to a scalar one, in which only the deformation with the same symmetry than the applied stress is considered. This of course greatly simplifies the problem, and speeds up the numerical computations necessary to calculate the material properties.

There is evidence that when the material that is being simulated is homogeneous the results using scalar models is quite accurate\cite{alexander1,vandem1}. This also includes the case in which the homogeneity of the material is obtained on average over a long period of time. For instance, it may happen that the local yield strength of the material at a fixed time has some spatial dependence. However, if in a very long time period the time average of the local yield strength is uniform, then the scalar implementation will produce very good results. 

In the present paper we want to analyze a case in which this average spatial homogeneity does not exist. It corresponds to a composite material, in which a host material which is homogeneous in the sense of the preceding paragraph, is added with inclusions of a second material with different properties. The inclusions may typically have different elastic or yielding properties than the host. 
In this case the inclusions keep their different elastic/plastic properties in time, and the time average argument does not apply. 
It could be argued that the composite is homogeneous if its properties are {\em spatially} averaged over distances larger than the typical distance among inclusions, which may be typically microscopic. However, it turns out that this does not imply that a scalar description of the global plasticity is accurate. 
In fact, we will show that only a full tensorial description of the problem provides sensible results in these cases.

In the next section we sketch the strategy to obtain the model equations in full tensorial description for a two dimensional system. We clearly indicate the approximations that should be made if we restrict to a scalar description. Then in Section III we present numerical simulations, and compare the results obtained using the scalar model with those of the tensorial description. 
In Sections IV and V we extend the results of Sections II and III to three dimensional samples. Finally, in Section VI we present our conclusions.

\section{General modeling of elasto/plastic properties in terms of the strain tensor}

Here we describe the simulation method for the two dimensional case.
The origin of the method can be traced back to the work of Bulatov and Argon\cite{bulatov}, and was then applied in quite different contexts \cite{r1,r2,r3,r4,r5}.
The starting point is to consider the (infinitesimal, or linearized) strain tensor $\varepsilon_{ij}$
in terms of the displacement field $u_{i}$

\begin{equation}
\varepsilon_{ij}=\frac 12 \left (\frac{\partial u_i}{\partial x_j}+\frac{\partial u_j}{\partial x_i}\right )
\end{equation}
where $i,j=1,2$. From here we define one volumetric
\begin{equation}
e_1\equiv (\varepsilon_{11}+\varepsilon_{22})/2
\end{equation}
and two deviatoric strains
\begin{eqnarray}
e_2&\equiv& (\varepsilon_{11}-\varepsilon_{22})/2\\
e_3&\equiv& \varepsilon_{12}
\end{eqnarray}
The deviatoric strains are related by a symmetry rotation of $45 \deg$.
The overdamped equations of motion are obtained by equating the time derivatives of $e_i$ to (minus) the variation of the total free energy $F$ with respect of $e_i$. However, in this process it has to be remembered that $e_1$, $e_2$, $e_3$ are not independent, but are related through
\begin{equation}
Q_1e_1+Q_2e_2+Q_3e_3=0
\label{stv2d}
\end{equation}
(with $Q_1\equiv\partial^2_x+\partial^2_y$, $Q_2\equiv\partial^2_y-\partial^2_x$, $Q_3\equiv -2\partial_x\partial_y$)
that follows immediately as an identity after writing $e_1$, $e_2$, $e_3$ in terms of $u_{ij}$. Thus using a Lagrange multiplier $\Lambda$ to satisfy the constraint, the equations of motion are written as
\begin{equation}
\lambda \dot e_i=f_i +\Lambda Q_i\\
\end{equation}
where $f_i\equiv -\frac{\delta F}{\delta e_i}$ define the local forces, and $\lambda$ is an overall effective viscosity coefficient.
To satisfy the compatibility constraint, we require
\begin{equation}
\Lambda =-\frac{\sum f_iQ_i}{2Q_4}
\label{lambda}
\end{equation}
with 
\begin{equation}
2Q_4\equiv Q_1^2+Q_2^2+Q_3^2=2(\partial_x^2+\partial_y^2)^2
\end{equation}

To write down the dynamical equations explicitly, we must specify the form of a free energy.
An isotropic elastic material es defined as having a free energy density given by
\begin{equation}
F_{el}=\int  \left (\frac B2 e_1^2+\frac{\mu}2 (e_2^2+e_3^2)\right )dxdy
\label{energia2d}
\end{equation}
The values of $B$ and $\mu$ (which can vary across the sample) are the local (two dimensional) bulk and shear modulus. 
To model a plastic material we must allow for the existence of plastic deviatoric strain, that we call $e_{20}$, $e_{30}$.
Note that these quantities do not satisfy any additional constraint. 
The free energy for given values of $e_{20}$, $e_{30}$ is then written as
\begin{equation}
F_{am}=\int  \left (\frac B2e_1^2+\frac{\mu}2 \left [(e_2-e_{20})^2+(e_3-e_{30})^2 \right]\right )dxdy
\end{equation}
and the equations of motion are explicitly written as
\begin{eqnarray}
\lambda \dot e_1&=&-Be_1+ Q_1\Lambda\label{e1}\\
\lambda \dot e_2&=&-\mu (e_2-e_{20})+ Q_2\Lambda\label{e2}\\
\lambda \dot e_3&=&-\mu (e_3-e_{30})+ Q_3\Lambda\label{e3}
\end{eqnarray}
and Eq. (\ref{lambda}) becomes

\begin{equation}
\Lambda =\frac{BQ_1e_1+\mu[Q_2(e_2-e_{20})+Q_3(e_3-e_{30})]}{2Q_4}
\label{lambda2}
\end{equation}
These are the dynamical equations of the system.

We will consider for simplicity a limiting case in which equations simplify a bit further. It corresponds
to a situation where the bulk modulus is taken to be much larger than the shear modulus, $B\gg \mu$.
In this case, Eqs. (\ref{e1}),(\ref{e2}),(\ref{e3}) show that $e_1$ has a much more rapid dynamics than $e_2$ and $e_3$,
and can be always considered to be at equilibrium, namely
\begin{equation}
e_1 =Q_1\Lambda/B
\end{equation}
Now eliminating $e_1$ from here and Eq. (\ref{lambda2}) we get
\begin{equation}
\Lambda =\frac{\mu[Q_2(e_2-e_{20})+Q_3(e_3-e_{30})]}{Q_4}
\end{equation}
The dynamical equations are then written explicitly in this case as 
\begin{eqnarray}
\lambda \dot e_2=f_2- \frac{Q_2^2}{Q_4}f_2 - \frac{Q_2Q_3}{Q_4}f_3  \label{2}\\
\lambda \dot e_3=f_3- \frac{Q_2Q_3}{Q_4}f_2- \frac{Q_3^2}{Q_4}f_3\label{3}
\end{eqnarray}
with $f_2=-\mu(e_2-e_{20})$, and $f_3=\mu(e_3-e_{30})$.
We typically perform the simulations calculating $f_2$, $f_3$ in real space, then Fourier transforming, 
and considering Eqs. (\ref{2}),(\ref{3}) in Fourier space. In this case, the form of the interaction kernels is

\begin{eqnarray}
\frac{Q_2^2}{Q_4}&=&\frac{(q_x^2-q_y^2)^2}{(q_x^2+q_y^2)^2}\label{eshelby}\\
\frac{Q_3^2}{Q_4}&=&\frac{4q_x^2q_y^2}{(q_x^2+q_y^2)^2}\label{eshelby2}\\
\frac{Q_2Q_3}{Q_4}&=&\frac{2q_xq_y(q_x^2-q_y^2)}{(q_x^2+q_y^2)^2}\label{eshelby3}
\end{eqnarray}

The previous equations leave the value of the uniform mode ${\bf q}=0$ undefined. Its evolution is fixed by the driving condition imposed. For instance, for a deformation at constant rate $\dot\gamma$ with the symmetry of $e_2$, the uniform mode is set as
\begin{eqnarray}
\overline{e_2}&=&\dot\gamma t \\
\overline{e_3}&=&0
\end{eqnarray}
where the bar indicates average on the whole system.

It remains to define the way in which the plastic strains $e_{20}$, $e_{30}$ evolve in time.
In the spirit of the Shear Transformation Zone theory\cite{stz}, we will consider that these quantities remain fixed when the local forces $f_2$, $f_3$ are sufficiently small. In this case the material behaves effectively as an elastic solid. 
However, when $f_2$, $f_3$ become too large, a plastic reacommodation (namely, a variation of $e_{20}$, $e_{30}$) occurs.
We use as a local yielding prescription the von Mises criterion, namely, reacommodation occurs when the elastic energy reaches some local threshold $\Omega$, i.e.,
\begin{equation}
(e_2-e_{20})^2+(e_3-e_{30})^2=2\Omega/\mu\equiv \kappa^2
\label{vm}
\end{equation}
The value of $\kappa$ is taken from a random distribution (in order to account for the disordered nature of an amorphous plastic material), and renewed every time condition \ref{vm} is met. The form of the distribution is the same for all sites of the host material, then making the host uniform on the long run.
Each time Eq. (\ref{vm}) is satisfied, we relax the elastic energy allowing $e_{20}$, $e_{30}$ to approach the current $e_2$, $e_3$. Although we could set $e_{20}$, $e_{30}$ to become the precise values of $e_2$, $e_3$ (relaxing elastic energy to zero), we will consider (taking into account the findings in molecular dynamic simulations by Nicolas et al. \cite{alexander3}) that the local plastic rearrangements are typically not exactly aligned with the axis of local maximum stress. Then our prescription will be to set new values of $e_{20}$, $e_{30}$ according to

\begin{eqnarray}
e_{20}^{new}=e_2+ \eta_2\label{eta2} \\
e_{30}^{new}=e_3+ \eta_3\label{eta3}
\end{eqnarray}
where $\eta_2$, $\eta_3$ are Gaussian variables with some width of the order of the typical value of $\kappa$. Note that the use of Gaussian random variables places the new values of $e_{20}$, $e_{30}$ isotropically around $e_2$, $e_3$, thus preserving the isotropy of the material being simulated.
As it can be seen, there is a good piece of freedom in the details of the dynamics of the plastic strain. We have tried a few variations and observed that the results obtained are rather insensitive to these details. In particular, the results to be shown remain unaltered if $\eta_2$, $\eta_3$ are set to zero, namely, if the plastic rearrangements occur exactly along the direction of maximum local stress.

It is now interesting to see how the use of a scalar model to describe plasticity can be justified starting from the full tensorial description we are presenting here. If the external driving is along one single deviatoric mode (let us suppose it to be $e_2$), then the scalar approximation appears if we assume that plastic behavior occurs only with this symmetry. In other words, in this case plastic deformation along $e_3$ is taken to be zero: $e_{30}\equiv 0$. Then Eq. (\ref{2}) becomes

\begin{equation}
\frac{\lambda}{\mu}\dot e_2=-(e_2-e_{20})+ \frac{Q_2^2}{Q_4}(e_2-e_{20})+ \frac{Q_2Q_3}{Q_4}e_3
\end{equation}
and here $e_3$ can be substituted by its expression from the compatibility condition, namely $e_3=-(Q_2/Q_3)e_2$, providing

\begin{equation}
\frac{\lambda}{\mu}\dot e_2=-(e_2-e_{20})- \frac{Q_2^2}{Q_4}e_{20}\label{escalar}
\end{equation}
Note that for this transformation to be valid, the value of the shear modulus must be one and the same all across the sample. 
This is the final expression. It represents a scalar model in which $e_2$ evolves in time due to the elastic force, and also due to the effect of the plastic strain all across the system, propagated through the kernel  $\frac{Q_2^2}{Q_4}$. Note that this kernel 
(Eq. (\ref{eshelby})) is nothing but the Eshelby interaction that is used in elasto-plastic models of plasticity. 
The dynamics of the plastic strain $e_{20}$ in the case of the scalar model is considered to be the following. $e_{20}$ is kept fixed as long as $|e_2-e_{20}|<\kappa$. When this threshold is reached, $e_{20}$ is renewed to a value $e_{20}^{new}=e_2+\eta$, with $\eta$ a stochastic variable of zero mean and a width or order of $\kappa$.
The connection between a scalar model like the one defined by Eq. (\ref{escalar}) and more standard implementation of elasto-plastic models has been recently elucidated in \cite{ferrero}.
We will now apply either the tensorial model (Eqs. (\ref{2}),(\ref{3})) or the scalar one (Eq. (\ref{escalar})) to describe the properties of a plastic composite and compare the results.

\section{Results for two-dimensional composites}
\label{section:2d}

We will present here the results for a macroscopically homogeneous two-dimensional sample (a square sample with periodic boundary conditions) driven by the application of a uniform deformation of symmetry $e_2$ at a constant rate $\dot \gamma$. 
The main interest will be in the average stress that appears in the sample under the action of this deformation rate. In particular, the stress when $\dot \gamma\to 0$ defines the overall yield stress of the sample $\sigma_c$. We will also pay special attention to the spatial distribution of the plastic deformation.

In the first place we consider the case of a pure host sample and compare the results obtained with the scalar and tensorial models. Particular details of the simulations are the following. System size is 128$\times$128, $\lambda/\mu=1$, and the value of $\kappa$ for the implementation of von Mises criterion (Eq. (\ref{vm})) is uniformly chosen between 1 and 2, and $\eta_2$, $\eta_3$ (Eqs. (\ref{eta2}),(\ref{eta3})) have a width of 0.2.

\begin{figure}
\includegraphics[width=9cm,clip=true]{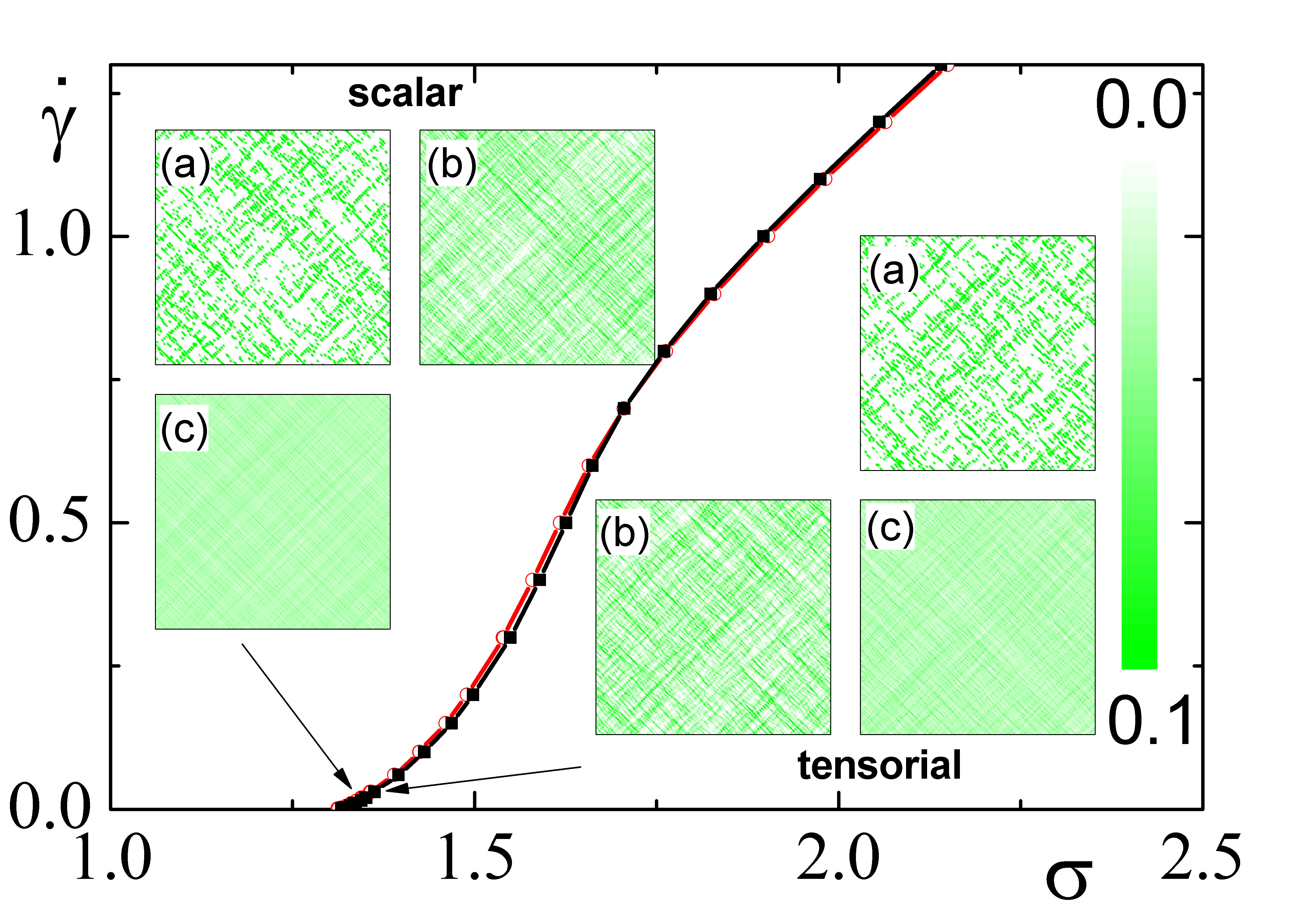}
\caption{The flow curve for a host without inclusions, calculated with the scalar (red, open circles), and the tensorial (black, full squares) model. There are only minor differences between the two curves. We also show the average plastic deformation rate in a simulation at $\dot\gamma=0.03$, during times intervals corresponding to a nominal global deformation of $0.3$(a), $3$(b), and $30$(c). The short time correlated plastic events distribute uniformly across the sample for large deformations.
\label{phi_eq_0}
}
\end{figure}

\begin{figure}
\includegraphics[width=8cm,clip=true]{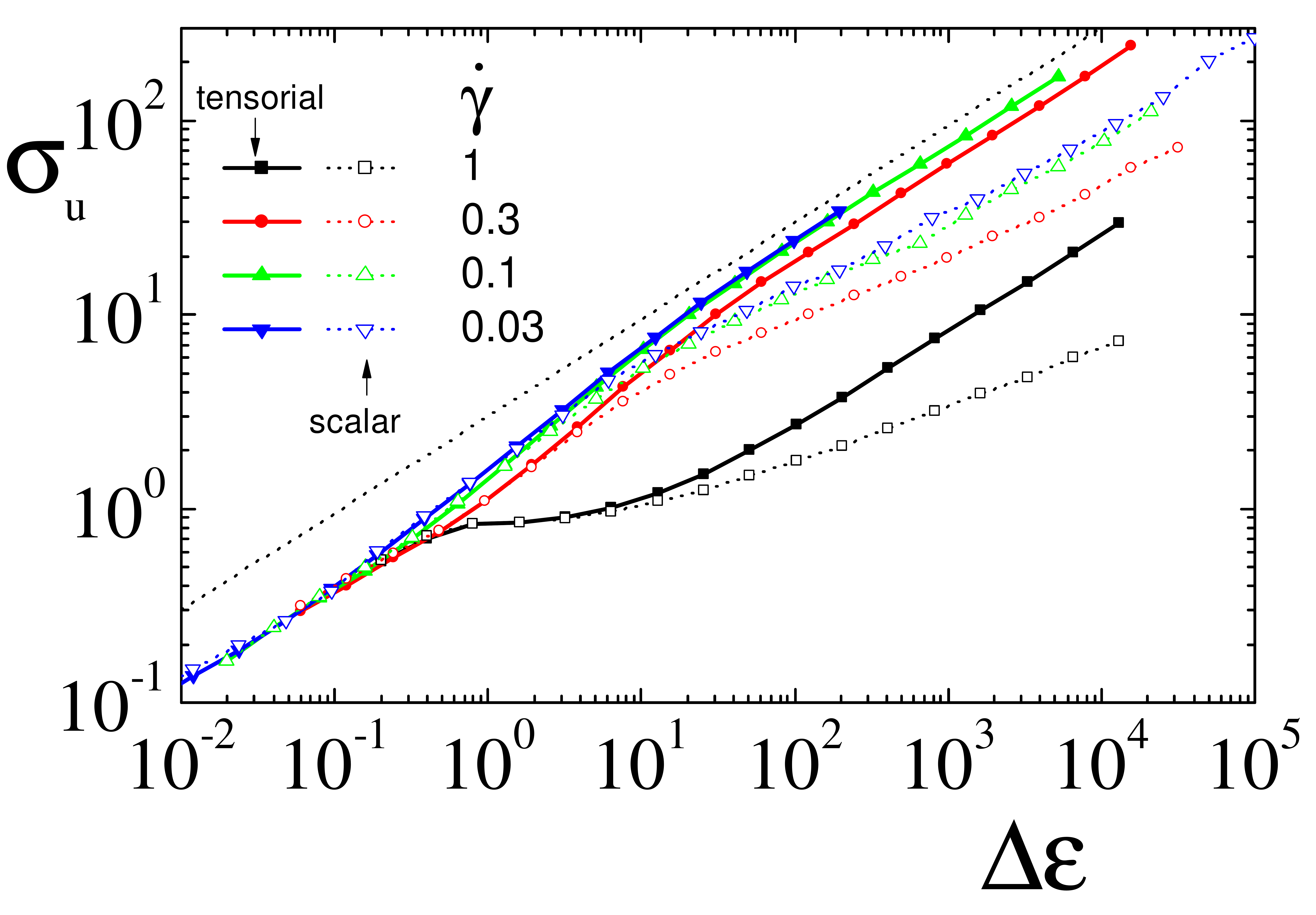}
\caption{Spread of the plastic deformation across the sample, as a function of the average deformation ($\Delta\varepsilon\equiv \dot\gamma \Delta t$) for different values of the external strain rate. Dotted line indicates the diffusive ($\sigma_u\sim \Delta \varepsilon^{1/2}$) behavior. 
\label{spread}
}
\end{figure}

The results obtained for the flow curve are contained in Fig. \ref{phi_eq_0}. Scalar and tensorial implementation provide practically equivalent results, as it was found already in a number of previous works\cite{alexander1,vandem1}. When looking in more detail at the spatial distribution of deformation (Fig. \ref{phi_eq_0}, insets), it is seen that in short time periods there is a strong tendency of the deformation to appear correlated along the diagonals of the sample
(that are the soft directions for the $e_2$ symmetry). However, when averaging the plastic deformation over large periods of time deformation becomes uniform. This behavior can be explicitly verified in the following way. Suppose that (for a given value of the applied $\dot \gamma$) we calculate the local average (over a time interval $\Delta t$) velocity of plastic deformation 

\begin{equation}
u({\bf r}, \Delta t)\equiv\frac{\Delta e_{20}({\bf r})}{\Delta t}
\end{equation}
at any position ${\bf r}$ of the sample. The spatial average of this quantity is simply $\dot \gamma$. By calculating its spread

\begin{equation}
\sigma_u(\Delta t)\equiv \sqrt{\overline{u^2}-\dot\gamma^2}
\end{equation}
we can tell if the deformation becomes uniform as $\Delta t\to \infty $, or if different regions have different 
average deformation rates. 
The results from the numerical simulation (Fig. \ref{spread}) clearly show a sub-linear behavior of $\sigma_u(\Delta t)$ as $\Delta t\to \infty$, namely, the plastic deformation is uniform in the long run. Note that the spread is lower (for the same total applied strain) at larger applied strain rates, but saturates to a well defined behavior as $\dot\gamma\to 0$. 
For the tensorial model, a typical diffusive behavior ($\sigma_u(\Delta t)\sim \Delta t^{1/2})$  is observed. In the scalar case,  the increase of $\sigma_u(\Delta t)$ seems to be slower, although it may be that we have not reached yet the diffusive regime.
The increase without limit of $\sigma_u(\Delta t)$ with $\Delta t$ (either in a diffusive or in a different way)
is a consequence of the zero modes of the interaction kernel. Should the kernel be strictly positive for any ${\bf q}\ne 0$ the amplitude of any mode should saturate, and so should the value of $\sigma_u(\Delta t)$ for large $\Delta t$. 
Note in this respect that the lower value of $\sigma_u$ in the scalar case can be associated to the underestimation of the number of zero modes in the scalar case, compared to the true tensorial one.

The main conclusion of the present case of a pure, uniform host is that the use of a scalar model captures essentially all important features of the more complete tensorial modeling.

\begin{figure}
\includegraphics[width=8cm,clip=true]{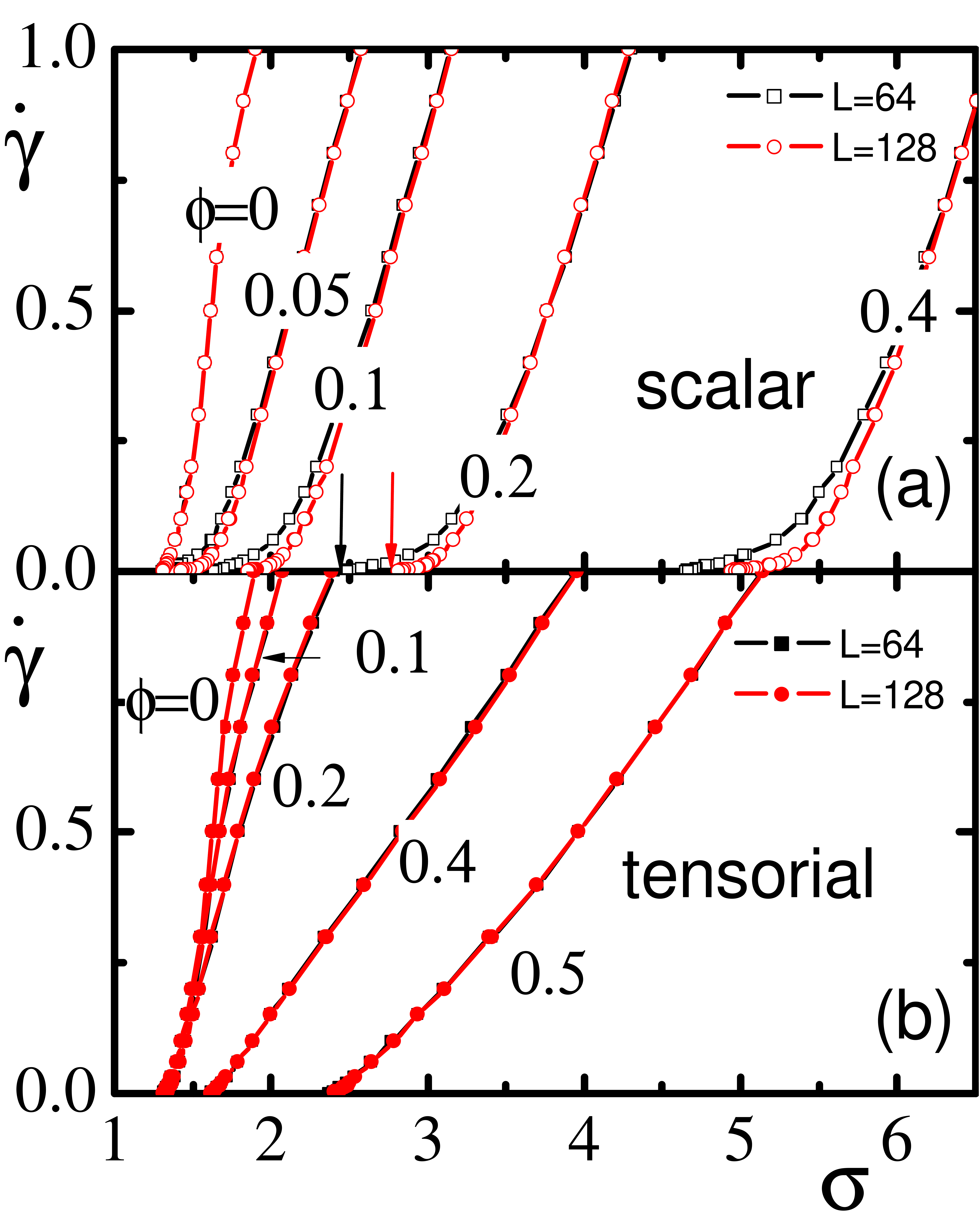}
\caption{Flow curves for different values of $\phi$ for systems of two different sizes ($L\times L$), using the scalar (a) and the tensorial model (b). Note the strong size effect in the scalar case at low values of $\dot\gamma$ (the size dependence of the yield stress $\sigma_c$ in the curve for $\phi=0.2$ is indicated), and the very different form of scalar and tensorial results when $\phi\ne 0$.
\label{fig_escalar}
}
\end{figure}

Things become different when a fraction $\phi$ of sites are replaced by inclusions with different elastic and/or plastic properties. The spatial distribution of inclusions is supposed to be totally uncorrelated, namely, each site of the sample has a probability $\phi$ of being an inclusion, and $(1-\phi)$ of being part of the host.
We consider the case in which the inclusions have the same elasticity (i.e., the same $\mu$) than the host, but a different plastic threshold. In concrete, for the inclusions the value of $\kappa$ (Eq. (\ref{vm})) is chosen to differ by a factor $h$ from that of the host. For $h>1$ this represents 
the case of harder inclusions, and some sort of ``hardening" effect of the whole material is expected.
This case was studied in \cite{vandembroucq} using a scalar model. The first thing to be shown here is that with our present implementation of the scalar model the results obtained reproduce those in \cite{vandembroucq}. 

\begin{figure}
\includegraphics[width=8cm,clip=true]{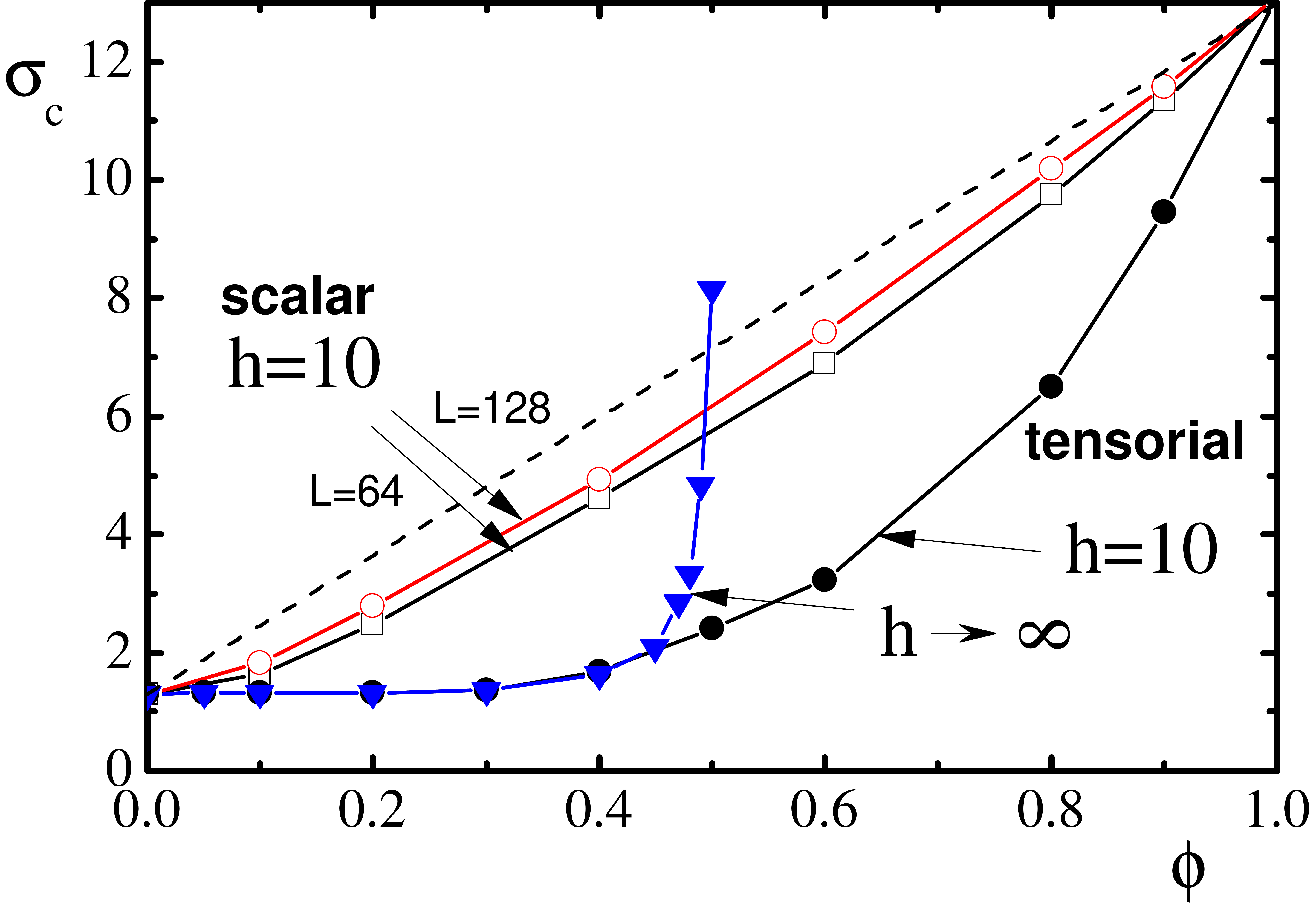}
\caption{Yield stress of a sample with a fraction $\phi$ of inclusions with a yield stress larger by a factor $h$ compared to the host. Results for the scalar model with $h=10$ with two system sizes to emphasize the strong size effect. In the tensorial case (with $L=128$) we show the results for $h=10$ (to be compared with the corresponding scalar case), and $h\to \infty$ (i.e., inclusions that do not yield at all).
\label{sigma_c}
}
\end{figure}

\begin{figure}
\includegraphics[width=8.5cm,clip=true]{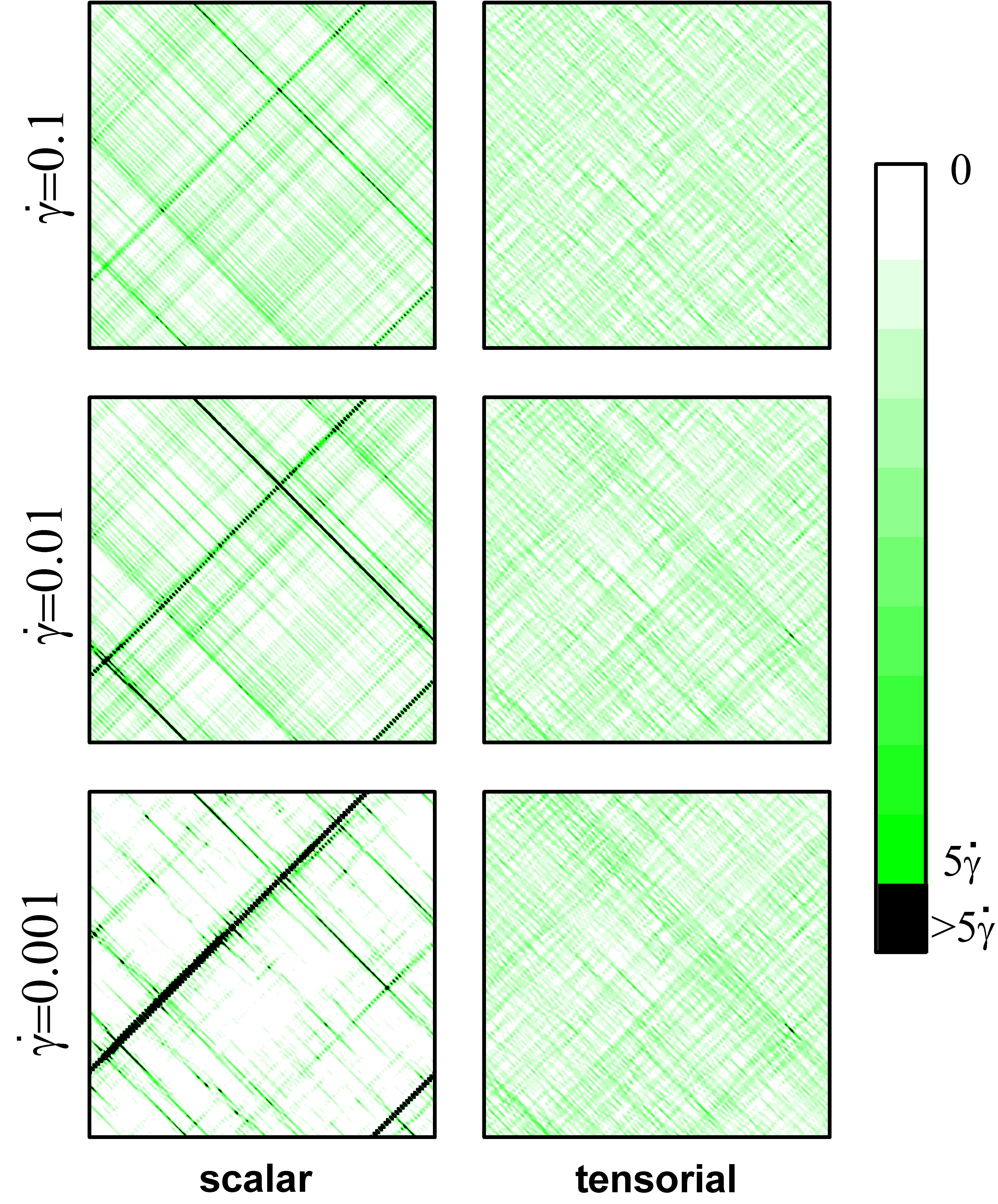}
\caption{Distribution of the plastic deformation for $\phi=0.2$ in very long runs at different values of $\dot\gamma$ as indicated, using the scalar (left images) and tensorial model (right images).
\label{localizacion}
}
\end{figure}

Fig. \ref{fig_escalar}(a) shows the flow curves for different values of $\phi$ and for two different system sizes, using the scalar model. As a general rule, the presence of harder inclusions shift the stress in the system to larger values, for a given value of $\dot\gamma$. Curves for different system sizes show a very strong size effect at low values of $\dot\gamma$, but a much weaker one at larger $\dot\gamma$.
If the yield stress $\sigma_c$ is plotted as a function of $\phi$ we obtain the results in Fig \ref{sigma_c}(open symbols). This qualitatively reproduce results in Fig. 6 of \cite{vandembroucq}.
There, the authors analyze the strong size effects and conclude that for $N\to \infty$ the value of the yield stress interpolates linearly between the value for the pure host $\sigma_H$ and that for pure inclusions $\sigma_I$, namely $\sigma_c(\phi)=\phi\sigma_H+(1-\phi)\sigma_I$.
The very strong size effects are attributed to the fact that inclusions tend to localize the deformation in certain regions of the sample and reduce it in others. 
In the scalar model the large $\Delta t$ behavior of the quantity 
$u({\bf r}, \Delta t)$ previously defined, has to be composed only by Fourier modes with zero energy.
The most generic function satisfying this fact is
\begin{equation}
u({\bf r}, \Delta t\to \infty)= u_1(x+y)+ u_2(x-y)
\end{equation}
for arbitrary functions $u_1$ and $u_2$.
The results confirm this analysis. Fig. \ref{localizacion} shows the accumulated plastic strain over a very large time interval, at decreasing values of $\dot\gamma$. For the scalar case (left images) the structure in terms of two different functions along the diagonals is clearly visible. In addition, the tendency to localize the deformation in a single narrow slip region is clearly visible as $\dot\gamma$ is reduced. The same trend occurs even for a single inclusion in a perfect host, no matter how large the hosts is. 
This is the origin of the very large size effect that is seen in Fig. \ref{fig_escalar}(a) and in Ref. \cite{vandembroucq}.

The situation changes dramatically when the full tensorial model is used. Fig. \ref{fig_escalar}(b) shows curves equivalent to those in (a), but the results are very different. First of all, the strong size effects of the scalar case have disappeared. Also the overall form of the curves and the values of $\sigma_c$ are quite different from those of the scalar case. The yield stress (Fig. \ref{sigma_c}) up to values of $\phi\simeq 0.3$ is almost identical to that for $\phi=0$. This behavior (that is also observed even in the case in which inclusions do not yield at all (Fig. \ref{sigma_c}, blue curve) is an indication that for low strain rates and not too large $\phi$, the system is able to accommodate the deformation without accumulating plastic deformation at the inclusions. This is now possible because the zero modes of the tensorial model are not limited to the two set of lines at 45 $\deg$, but consists of any straight line at arbitrary angle. This becomes apparent by comparing the spatial distribution of plastic deformation in the scalar  case (Fig. \ref{localizacion} left panels), with the present tensorial case (Fig. \ref{localizacion} right panels). In the scalar case plastic deformation is more and more localized as $\dot\gamma\to 0$, as discussed in \cite{vandembroucq}. However we see in the more realistic tensorial modeling that plastic deformation remains finite in a a finite fraction of the sample even in the limit $\dot\gamma\to 0$. This may be considered the fundamental difference between scalar and tensorial modeling, and responsible for all differences we are observing in the simulations.

\begin{figure}
\includegraphics[width=8cm,clip=true]{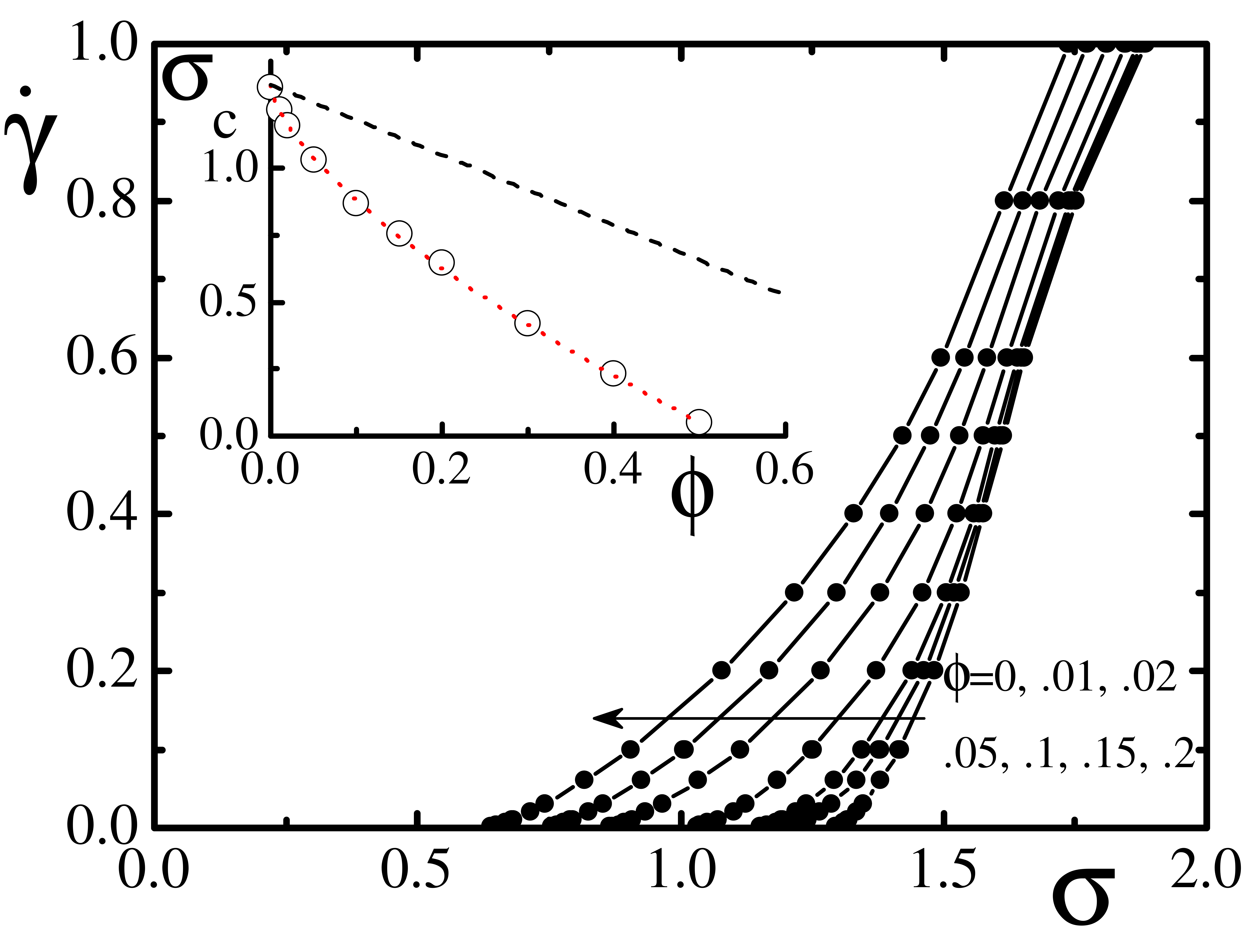}
\caption{Flowcurves in the presence of a fraction $\phi$ of ``liquid bubbles" in the system (i.e., regions with $\mu=0$) ($L=128$), using the tensorial model. The inset shows the dependence of the yield stress with $\phi$. Red dotted line shows the function $\sigma(\phi)=\sigma(0)-C\phi^{2/3}$. Black 
dashed line is the ``linear interpolation" result.}
\label{bubbles}
\end{figure}

A situation that does not seem to have been previously investigated is the possibility of having inclusions that are softer than the host. It is expected than these inclusions produce an overall softer composite, but it is not clear to what extent.
We model a limiting case of softer inclusions, considering them as spots where $\mu=0$ (or alternatively, where the local yield stress is zero). This is like considering that we are introducing in the host a certain amount of ``liquid bubbles". 
The results of the flowcurves in this case can be seen in Fig. \ref{bubbles} (we only show curves for the realistic tensorial modeling in this case).
 Although for large values of $\dot\gamma$ the behavior of the flow curves seems to be simply shifted downward in an amount proportional to $\phi$, at very low values of $\dot\gamma$ the decreasing is more dramatic. We associate this effect to the fact that at very low strain rates, regions of plastic deformation can snake through the impurities finding softer paths of deformations, then reducing $\sigma_c$ in a larger extent. The plot of $\sigma_c$ vs. $\phi$ (Fig. \ref{bubbles}, inset) indicates that at very low values of $\phi$ the behavior seems to be super-linear. In fact, we fit $\sigma_c(\phi)=\sigma_c(0)-C \phi^{2/3}$ pointing to non-trivial long range interactions among the inclusions. 

\section{Modeling of three-dimensional systems}

The previous results for two-dimensional systems clearly point to the necessity of being careful when deciding if a scalar modeling of elasto-plastic behavior can produce sensible results or not. We have made clear that in order to understand the behavior of 2D composites a full tensorial modeling is necessary. 

We should expect the same behavior in 3D. Modeling 3D systems can be proposed along the same lines used for 2D systems, with the expected increase of algebraic complexity. We sketch the main steps here, and refer to the Appendix for the full expressions. In 3D we start from the $3\times 3$ symmetric strain tensor that contains one volumetric (noted $e_1$) and 5 deviatoric ($e_2$, ... $e_6$) strains. The deviatoric modes can be defined in such a way that the free energy of a perfect elastic isotropic material is written (compare with Eq. (\ref{energia2d})) as

\begin{equation}
F_{el}=\int  \left (\frac B2 e_1^2+\frac \mu  2 \sum_{j=2}^6 e_j^2\right )d^ 3r
\end{equation}
In order to model amorphous plastic systems, plastic strain fields $e_{20}$,...,$e_{60}$ are introduced in such a way that
\begin{equation}
F_{am}=\int  \left (\frac B2 e_1^2+\frac \mu 2 \sum_{j=2}^6 (e_j-e_{j0})^2 \right )d^ 3r
\end{equation}

There are 3 compatibility constrains in 3D (that reduce the 6 variables $e_1$... $e_6$ to 3 true degrees of freedom) that can be considered to be equations of the form of Eq. (\ref{stv2d}) for planes $xy$, $yz$, and $zx$. After a rather long process (sketched in the Appendix) that involves introducing three Lagrange multipliers, and taking already the limit $B\gg \mu$, the dynamic equations of the system can be written as

\begin{equation}
\lambda \dot e_i=f_i+\sum_{j=2}^6 Q_{ij}f_j
\label{eq_3d}
\end{equation}
where $f_j\equiv -\mu (e_j-e_{j0})$, 
and $Q_{ij}$ are differential operators (or algebraic operators in Fourier space) analogous to those defined in Eqs. \ref{eshelby},\ref{eshelby2},\ref{eshelby3}. The explicit form of $Q_{ij}$ is presented in the Appendix.

The modeling of plastic strain is made in analogy with 2D, by calculating the deviatoric elastic energy $E_{dev}$ 
\begin{equation}
E_{dev}=\frac \mu 2 \sum_{j=2}^6 (e_j-e_{j0})^2 
\end{equation}
If $E_{dev}$ overpasses a von Mises threshold $\Omega$, new values of $e_{e0}$...$e_{60}$ are chosen according to 

\begin{equation}
e_{j0}=e_j+ \eta_j   
\end{equation}
where $\eta_j$ are random Gaussian variables (to preserve the rotational symmetry in the $e_2$, ..., $e_6$ space).

As in the 2D case, if all but one (this is supposed to be $e_2$) deviatoric stresses are assumed to be perfectly elastic (i.e., $e_{j0}=0$ for $j=3,4,5,6$) then the explicit forms of $e_3$...$e_6$ can be plugged into the equation for $e_2$, to produce  the equation 
\begin{equation}
\lambda \dot e_2=f_2+\mu Q_{22}e_{20}
\end{equation}
This is the way in which a ``scalar" model in the 3D case appears.

\section{Results for three-dimensional composites}

\begin{figure}
\includegraphics[width=8cm,clip=true]{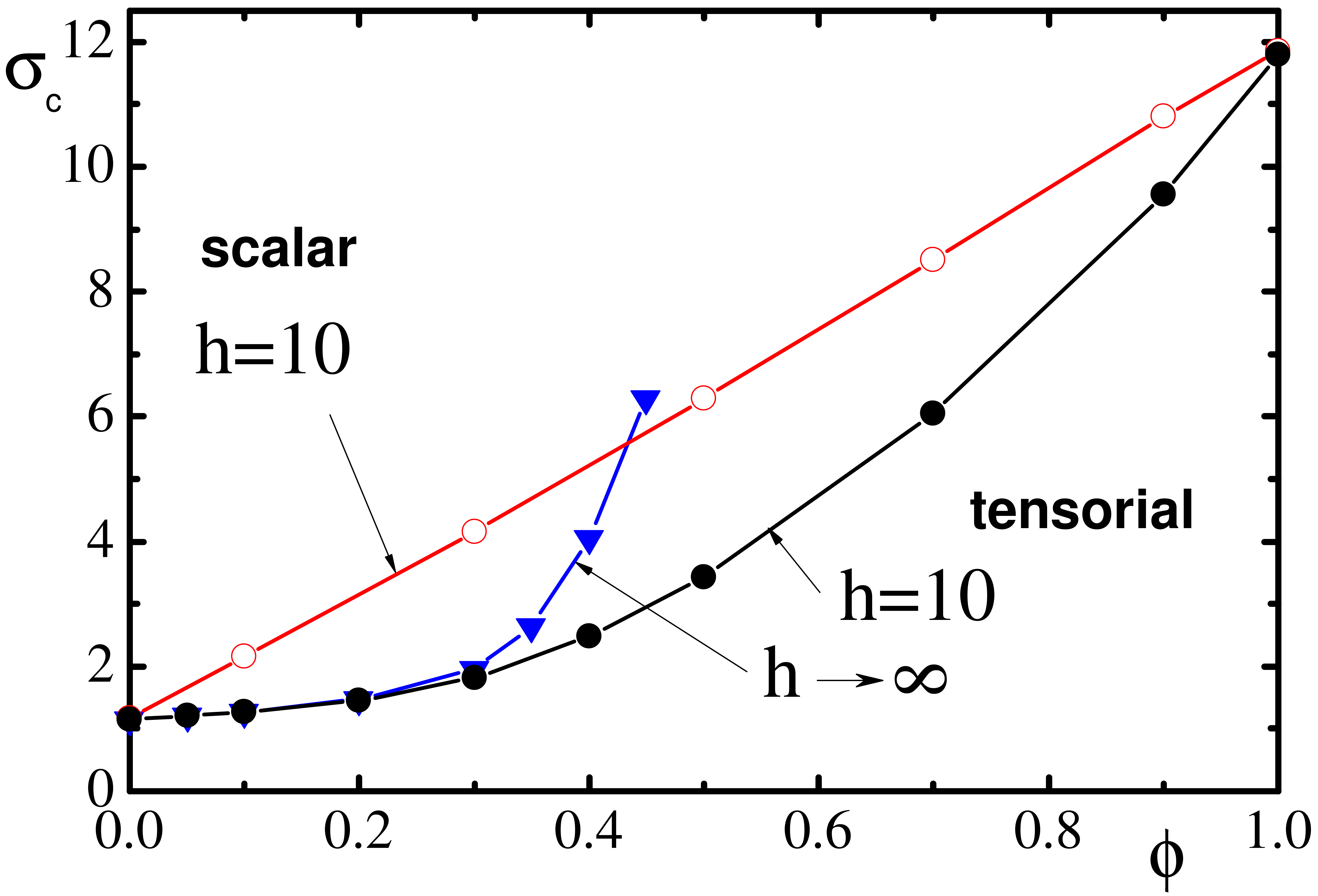}
\caption{Yield stress as a function of the fraction of inclusions $\phi$ (which have a yielding threshold with a factor of $h$ different than the host) in three dimensions. Now the scalar modeling does not display large size effect, and all results are presented for $L=24$. The linear interpolation behavior displayed by the scalar model does not match the results obtained using the tensorial model.
\label{sigma_c_3d}
}
\end{figure}

There is a twofold  interest in generalizing the results in Section \ref{section:2d} to three dimensions.
On one side, the finding that in 2D scalar and tensorial modeling give very different results raises the expectation that the same occurs in 3D.
The second reason is that some results obtained in 2D using tensorial modeling are rather detrimental from a practical point of view, and it is important to see if they persist in 3D. For instance, in 2D we have obtained that a fraction of harder inclusions (up to approximately $\phi\simeq 0.3$) does not really produce any hardening effect on the material, i.e., does not increase its yield stress. It is important to see if this effect also occurs in 3D. 

\begin{figure}
\includegraphics[width=8cm,clip=true]{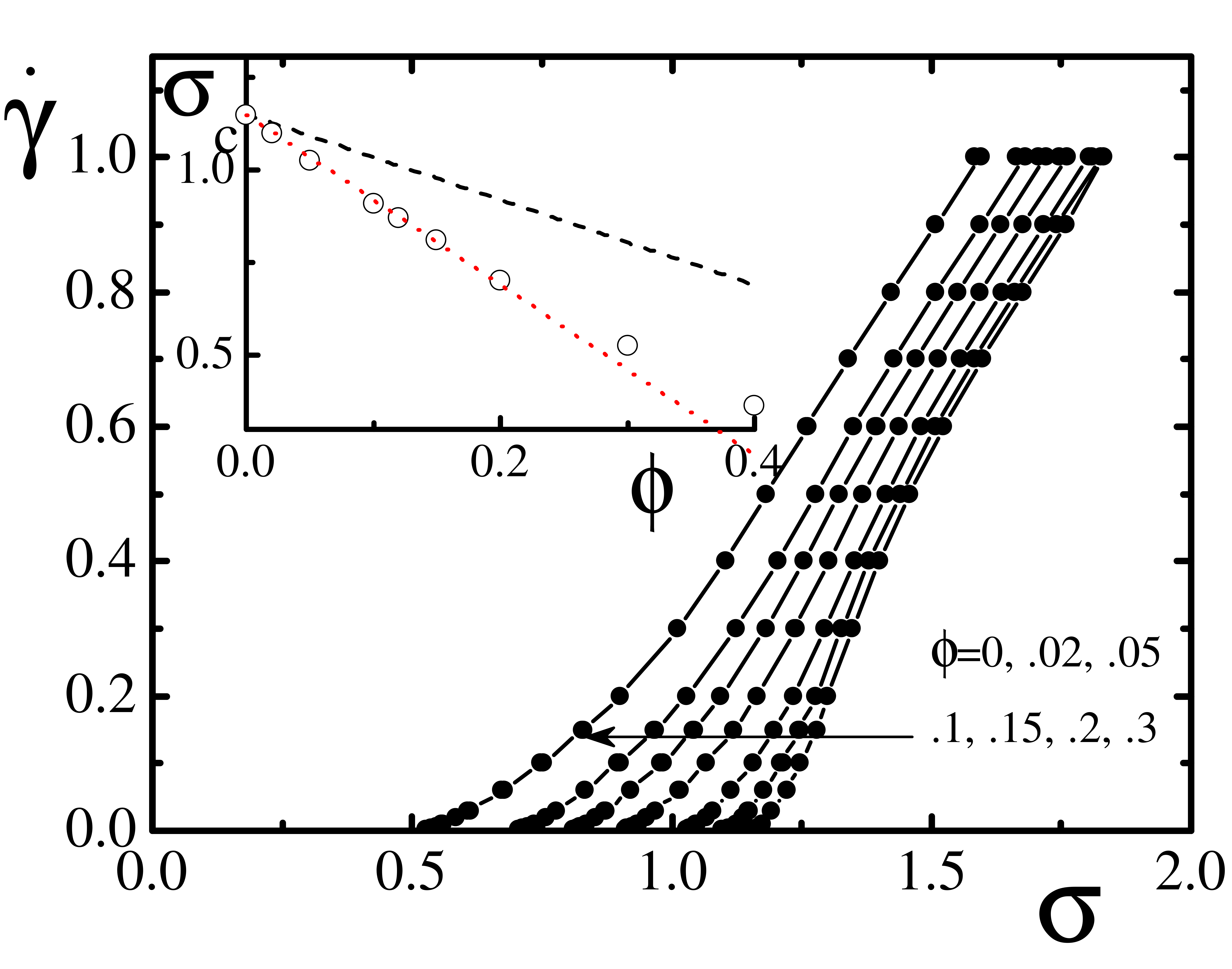}
\caption{Same as Fig. \ref{bubbles} for three dimensional systems ($L=24$). The decrease of $\sigma_c$ with $\phi$ is now linear, but about a factor of two more rapid than the linear interpolation result (black dashed line).
}
\label{bubbles_3d}
\end{figure}

The main results we will present for 3D systems are contained in Fig. \ref{sigma_c_3d}, which should be compared with the equivalent 2D case (Fig. \ref{sigma_c}). First of all, we mention that in scalar 3D modeling we have not observed the strong size effects that were present in 2D. Systems of size $L\gtrsim 16$ are already representative of the ``thermodynamic limit". Yet, the results for $\sigma_c$, that show an almost perfect linear increase of $\sigma_c$ with $\phi$ do not coincide with the results obtained using the more complete tensorial modeling. In fact, the full tensorial simulation shows that the effect of harder impurities is weak at low concentrations, yet much more appreciable than in 2D: we observe in 3D an approximately quadratic increase of $\sigma_c$ with $\phi$ for low values of $\phi$. This behavior persists even if the inclusions do not yield at all. 

The case of very soft inclusions displays also some interesting differences compared with the 2D case. Fig. \ref{bubbles_3d} shows the flow curves and the values of $\sigma_c$. Now the decrease of $\sigma_c$ with $\phi$ is linear at small values of $\phi$, yet it is about a factor of two larger than the one expected from the weighted average of yield stress of host and inclusions.

\section{Conclusions}
In this paper we have presented a full tensorial scheme to describe the elasto/plastic properties of materials (both in 2D and 3D) that accounts for the elastic compatibility of the material, and also for the possibility of plastic yielding. 
The model has been presented for the case in which the bulk modulus is much larger than the shear modulus, but the general case can be treated along the same lines in a straightforward way. As a side result, the conditions under which this full tensorial model reduces to a scalar one have been elucidated: to obtain a scalar model involving a single shear mode of definite symmetry, all remaining shear modes must behave harmonically (i.e, yielding must not occur in the remaining modes), and the shear coefficient of all remaining modes must be a single unique number, namely it is not allowed to fluctuate neither spatially, nor temporally. 

In addition, we have addressed in a particularly important case (namely that of a plastic material that contains harder, or softer inclusions) the differences that appear between the full (more realistic) tensorial modeling, and an approximate scalar modeling. Our result--consistent with previous findings--suggest that in the absence of inclusions, the use of a scalar model (that disregards yielding in the remaining shear mode) is quantitatively very precise to describe for instance the flow curve. Other works have shown that it is also appropriate to describe quantitatively the avalanche statistics in the system\cite{zapperi}.
In the case in which the material contains harder or softer inclusions, we have shown clearly that the results using a full tensorial model differ dramatically from those obtained using the scalar model. In particular, we do not find strong size effects as they were obtained using the scalar model, and the inclusion of a fraction up  to about 20 \% of harder inclusions does not affect appreciable the value of the shear stress of the material (even in the case in which the inclusions are infinitely hard) because plastic deformation can appear in the system along wavy paths that avoid the hard inclusions. It has to be kept in mind however, that these results apply to the present case of inclusions added to the host in a totally uncorrelated manner. In cases in which the inclussions are accommodated in particularly convenient forms, the hardening effect can be much more appreciable.

One of the main conclusions of the present work is that one has to be careful when deciding if a scalar model is enough to describe the elasto/plastic properties of a material. Even cases in which the system is uniform on average (as the present case) may require the use of a full tensorial machinery. We note that the explicit simulation scheme we have presented (particularly the formulae for the 3D case shown in the appendix) can have interesting applications to other problems such as the study of cases in which the sample is anisotropic and the orientation of the anisotropy varies from point to point, or cases in which there is the possibility to generate cracks in the sample (modeled for instance as regions where the bulk and shear modulus of the material are set down to zero) when some deformation threshold criterion is met.

\begin{widetext}

\section{Appendix}

Here we provide a more detailed derivation and the final expressions necessary to implement the full three dimensional modeling (Eq. (\ref{eq_3d})).
The starting point is the definition of the strain tensor $\varepsilon_{ij}$ in terms of the displacement field $u_{i}$

\begin{equation}
\varepsilon_{ij}\equiv \frac 12\left(  \frac{\partial u_i}{\partial x_j}+\frac{\partial u_j}{\partial x_i}   \right)
\end{equation}
From its very definition, components of $\varepsilon_{ij}$ satisfy identically the three constraints:

\begin{eqnarray}
(\partial_x^2+\partial_y^2)(\varepsilon_{11}+\varepsilon_{22})-(\partial_x^2-\partial_y^2)(\varepsilon_{11}-\varepsilon_{22})-4\partial_x \partial_y \varepsilon_{12}=0\\
(\partial_y^2+\partial_z^2)(\varepsilon_{22}+\varepsilon_{33})-(\partial_y^2-\partial_z^2)(\varepsilon_{22}-\varepsilon_{33})-4\partial_y \partial_z \varepsilon_{23}=0\\
(\partial_z^2+\partial_x^2)(\varepsilon_{33}+\varepsilon_{11})-(\partial_z^2-\partial_x^2)(\varepsilon_{33}-\varepsilon_{11})-4\partial_z \partial_x \varepsilon_{31}=0
\end{eqnarray}
Now, we define the volumetric
\begin{equation}
e_1\equiv \varepsilon_{11}+\varepsilon_{22}+\varepsilon_{33}
\end{equation}
and deviatoric strains 
\begin{eqnarray}
e_2&\equiv& (\varepsilon_{11}-\varepsilon_{22})/\sqrt 2\\
e_3&\equiv& (\varepsilon_{11}+\varepsilon_{22}-2\varepsilon_{33})/\sqrt 6\\
e_4&\equiv& \sqrt 2 \varepsilon_{12}\\
e_5&\equiv& \sqrt 2 \varepsilon_{23}\\
e_6&\equiv& \sqrt 2 \varepsilon_{31}.
\end{eqnarray}

In terms of deviatoric and volumetric strains, the constraints are written as 

\begin{eqnarray}
P_1e_1+Ae_2+De_3+Ge_4=0\nonumber\\
P_2e_1+Be_2+Ee_3+He_5=0\nonumber\\
P_3e_1+Ce_2+Fe_3+Ie_6=0\nonumber
\end{eqnarray}
where 
\begin{eqnarray}
P_1=\frac 23(\partial_x^2+\partial_y^2), ~~~~P_2=\frac 23(\partial_y^2+\partial_z^2), ~~~~P_3=\frac 23(\partial_z^2+\partial_x^2)\\
A=-\sqrt 2(\partial_x^2-\partial_y^2), ~~~~B=-\sqrt 2 \partial_z^2, ~~~~C=\sqrt 2 \partial_z^2\\
D=\frac{\sqrt 6}{3}(\partial_x^2+\partial_y^2), ~~~~E=\frac{\sqrt 6}{3}(\partial_z^2-2 \partial_y^2) , ~~~~F=\frac{\sqrt 6}{3}(\partial_z^2-2 \partial_x^2)\\
G=-2\sqrt 2 \partial_x\partial_y,~~~~H=-2\sqrt 2 \partial_y\partial_z,~~~~I=-2\sqrt 2 \partial_z\partial_x
\end{eqnarray}

We assume a free energy of the system of the form
\begin{equation}
F=\int d^3 r\left (\frac {B_0}2 e_1^2+ V(e_2,...,e_6)\right ),
\label{f}
\end{equation}
and overdamped equations of motion
\begin{eqnarray}
\lambda\dot e_1&=& -B_0e_1 -P_1\Lambda_1 -P_2\Lambda_2 -P_2\Lambda_3\\
\lambda\dot e_2&=& f_2 -A\Lambda_1 -B\Lambda_2 -C\Lambda_3 \\
\lambda\dot e_3&=& f_3 -D\Lambda_1 -E\Lambda_2 -F\Lambda_3 \\
\lambda\dot e_4&=& f_4 -G\Lambda_1 \\
\lambda\dot e_5&=& f_5 -H\Lambda_2 \\
\lambda\dot e_6&=& f_6 -I\Lambda_3 
\end{eqnarray}
where Lagrange multipliers $\Lambda_1$, $\Lambda_2$, $\Lambda_3$ are used to enforce the constraints, and $f_i=-\partial V/\partial e_i$ ($i=2$, ... 6). 

From now on, we will consider the bulk modulus $B_0$ to be very large compared with shear moduli in the system, in such a way that $e_1$ can be safely set to 0.
In this limit, by transforming to Fourier space, a direct but lengthy calculation yields

\begin{eqnarray}
\lambda \dot e_2&=&f_2
-\frac{(q_x^2-q_y^2)^2+q^2q_z^2}{q^4}f_2
+\frac{\sqrt{3}q_z^2(q_x^2-q_y^2)}{q^4}f_3
+\frac{2q_xq_y(q_y^2-q_x^2)}{q^4}f_4
-\frac{q_yq_z(3q_x^2-q_y^2+q_z^2)}{q^4}f_5
-\frac{q_xq_z(q_x^2-3q_y^2-q_z^2)}{q^4}f_6\nonumber\\
\lambda \dot e_3&=&f_3
+\frac{\sqrt{3}q_z^2(q_x^2-q_y^2)}{q^4}f_2
-\frac{q^4-3q_z^2(q_x^2+q_y^2)}{q^4}f_3
+\frac{2\sqrt{3}q_xq_yq_z^2}{q^4}f_4
-\frac{\sqrt{3}q_yq_z(q_x^2+q_y^2-q_z^2)}{q^4}f_5
-\frac{\sqrt{3}q_xq_z(q_x^2+q_y^2-q_z^2)}{q^4}f_6\nonumber\\
\lambda \dot e_4&=&f_4
+\frac{2q_xq_y(q_y^2-q_x^2)}{q^4}f_2
+\frac{2\sqrt{3}q_xq_yq_z^2}{q^4}f_3
-\frac{4q_x^2q_y^2+q_z^2q^2}{q^4}f_4
-\frac{q_xq_z(4q_y^2-q^2)}{q^4}f_5
-\frac{q_yq_z(4q_x^2-q^2)}{q^4}f_6\nonumber\\
\lambda \dot e_5&=&f_5
-\frac{q_yq_z(3q_x^2-q_y^2+q_z^2)}{q^4}f_2
-\frac{\sqrt{3}q_yq_z(q_x^2+q_y^2-q_z^2)}{q^4}f_3
-\frac{q_xq_z(4q_y^2-q_r^2)}{q^4}f_4
-\frac{4q_y^2q_z^2+q_x^2q^2}{q^4}f_5
-\frac{q_xq_y(4q_z^2-q^2)}{q^4}f_6\nonumber\\
\lambda \dot e_6&=&f_6
-\frac{q_xq_z(q_x^2-3q_y^2-q_z^2)}{q^4}f_2
-\frac{\sqrt{3}q_xq_z(q_x^2+q_y^2-q_z^2)}{q^4}f_3
-\frac{q_yq_z(4q_x^2-q^2)}{q^4}f_4
-\frac{q_xq_y(4q_z^2-q^2)}{q^4}f_5
-\frac{4q_z^2q_x^2+q_y^2q^2}{q^4}f_6\nonumber
\end{eqnarray}
where $q^2\equiv q_x^2+q_y^2+q_z^2$, $q^4\equiv (q^2)^2$, and $f_i$ must be understood as evaluated at the corresponding value of $\bf q$. These equations 
can be written as
\begin{equation}
\lambda \dot e_i=f_i+\sum_{j=2}^6 Q_{ij}f_j
\label{uf}
\end{equation}
that allow to define the $Q_{ij}$ used in Eq. (\ref{eq_3d}). 

Up to here, the model equations are given in terms of the total strain $e_i$, and the generalized forces $f_i\equiv -\partial V/\partial e_i$ obtained from a general free energy (Eq. (\ref{f})). In the case in which the form of the $V$ function is piece-wise parabolic, the forces become $f_i\equiv -\mu(e_i-e_{i0})$, where $e_{i0}$ can be identified with the ``plastic" strain. In this case, the equations simplify by noticing that all parts of $f_i$ proportional to $e_i$ in the last term of \ref{uf} sum up to zero, 
providing 
\begin{equation}
\lambda \dot e_i=f_i+\mu\sum_{j=2}^6 Q_{ij}e_{j0}
\end{equation}
From this form of the equations the scalar model is obtained very easily as the equation for $e_2$ if we assume that plastic deformation occurs only with the symmetry of $e_2$ and then $e_{j0}=0$ for $j=3$,...,6, namely

\begin{equation}
\lambda \dot e_2=f_2+\mu Q_{22}e_{20}
\end{equation}



\end {widetext}


\begin{thebibliography}{1}

\bibitem{torquato}S. Torquato, {\em Random Heterogeneous Materials. Microstructure
and Macroscopic Properties} (Springer, New York, 2002).

\bibitem{vandem_review} D. Rodney, A. Tanguy, and D. Vandembroucq,
{\em Modeling the mechanics of amorphous solids at different length and time scales}
Modell. Simul. Mater. Sci. Eng. {\bf 19}, 083001 (2011).

\bibitem{rmp2}D. Bonn, M. M. Denn, L. Berthier, T. Divoux and S. Manneville, 
{\em Yield stress materials in soft condensed matter},
Rev. Mod. Phys.,  {\bf 89}, 035005 (2017).

\bibitem{rmp}A. Nicolas, E. E. Ferrero, K. Martens, and J.-L. Barrat,
{\em Deformation and flow of amorphous solids: Insights from elastoplastic models},
Rev. Mod. Phys. {\bf 90}, 045006 (2018).

\bibitem{talamali} M. Talamali, V. Pet\"aj\"a, D. Vandembroucq, and S. Roux,
{\em Avalanches, precursors, and finite-size fluctuations in a mesoscopic model of amorphous plasticity},
Physical Review E {\bf 84} 016115 (2011).

\bibitem{pnas} J. Lin, E. Lerner, A. Rosso, and M. Wyart, 
{\em Scaling description of the yielding transition in soft amorphous solids at zero temperature},
Proceedings of the National Academy of Sciences {\bf 111}, 14382 (2014).

\bibitem{zapperi} Z. Budrikis, D. F. Castellanos, S. Sandfeld, M. Zaiser,
and S. Zapperi, 
{\em Universal features of amorphous plasticity},
Nat. Comm. {\bf 8}, 15928 (2017).

\bibitem{aguirre}I. Fern\'andez Aguirre and E. A. Jagla, 
{\em Critical exponents of the yielding transition of amorphous solids},
Phys. Rev. E {\bf 98}, 013002 (2018).

\bibitem{vandembroucq}B. Tyukodi, C. A. Lemarchand, J. S. Hansen, and D. Vandembroucq, 
{\it Finite size effects in a model for plasticity of amorphous composites},
Phys. Rev. E {\bf 93}, 023004 (2016).

\bibitem{indent} Z. Budrikis, D. Fernandez-Castellanos, S. Sandfeld, M. Zaiser, S. Zapperi,
{\em Universality of Avalanche Exponents in Plastic Deformation of Disordered Solids},
Nat. Comm. {\bf 8}, 15928 (2017)

\bibitem{alexander1}A. Nicolas, J. Rottler, and J. L.  Barrat,
{\em Spatiotemporal correlations between plastic events in the shear flow of athermal amorphous solids},
Eur. Phys. J. E {\bf 37}, 50 (2014).

\bibitem{vandem1}M. Talamali, V. Pet\"aj\"a, D. Vandembroucq, and S. Roux,
{\em Strain localization and anisotropic correlations in a mesoscopic model of amorphous plasticity},
Comptes Rendus M\'ecanique {\bf 340}, 275 (2012)

\bibitem{bulatov}V. V. Bulatov and A. S. Argon, 
{\em A stochastic model for continuum elasto-plastic behavior: I. Numerical approach and strain localization} 
Modell. Simul. Mater. Sci. Eng.{\bf 2}, 167 (1994).

\bibitem{r1} T.Lookman, S.R. Shenoy, K. O. Rasmussen, A. Saxena, and A.R. Bishop, 
{\em Ferroelastic dynamics and strain compatibility},
Phys. Rev. B {\bf 67}, 024114 (2003). 

\bibitem{r2} S. Kartha, J. A.Krumhansl, J. P. Sethna, and L. K. Wickham, 
{\em Disorder-driven pretransitional tweed pattern in martensitic transformations},
Phys. Rev. B {\bf 52}, 803 (1995).

\bibitem{r3} E. A. Jagla, 
{\em Maturation of crack patterns},
Phys. Rev. E {\bf 69}, 056212 (2004).

\bibitem{r4} E. A. Jagla,
{\em Morphologies of expansion ridges of elastic thin films onto a substrate},
Phys. Rev. E {\bf 74}, 036207 (2006).

\bibitem{r5}E. A. Jagla,
{\em Strain localization driven by structural relaxation in sheared amorphous solids},
Phys. Rev. E {\bf 76}, 046119 (2007).

\bibitem{stz}M. L. Falk and J. S. Langer,
{\em Dynamics of viscoplastic deformation in amorphous solids},
Phys. Rev. E {\bf 57}, 7192 (1998).

\bibitem{alexander3}A. Nicolas, and J. Rottler,
{\it Orientation of plastic rearrangements in two-dimensional model glasses under shear},
Phys. Rev. E {\bf 97}, 063002 (2018).

\bibitem{ferrero}E.E. Ferrero and E. A. Jagla,
{\em Criticality in elastoplastic models of amorphous solids with stress-dependent yielding rates},
Soft Matter, {\bf 15}, 9041 (2019).

\end{thebibliography}
\end{document}